\documentclass[conference]{IEEEtran}
\IEEEoverridecommandlockouts
\usepackage[utf8]{inputenc}
\usepackage[T1]{fontenc}
\usepackage{cite}
\usepackage{amsmath,amssymb,amsfonts}
\usepackage{graphicx}
\usepackage{textcomp}
\usepackage{xcolor}
\usepackage{url}
\usepackage[hidelinks]{hyperref}
\usepackage{balance}

\usepackage{eso-pic} % 
\begin{document}

\title{From Idea to Classroom in Days: Using ``Vibe Coding'' to Create a Programming Process Visualizer from IDE Activity Logs}

\author{\IEEEauthorblockN{Heidi Taveter}
\IEEEauthorblockA{\textit{Institute of Computer Science} \\
\textit{University of Tartu}\\
Tartu, Estonia \\
ORCID: 0000-0002-2638-7253}
\and
\IEEEauthorblockN{Marina Lepp}
\IEEEauthorblockA{\textit{Institute of Computer Science} \\
\textit{University of Tartu}\\
Tartu, Estonia \\
ORCID: 0000-0003-3303-5245}}

\maketitle
\AddToShipoutPictureFG*{%
  \AtPageLowerLeft{%
    \raisebox{1.8cm}{%
      \makebox[\paperwidth][c]{%
        \footnotesize
        This work has been accepted for presentation at the FIE 2026 conference and will be published in the IEEE proceedings.
      }%
    }%
  }%
}

\AddToShipoutPictureFG*{%
  \AtPageLowerLeft{%
    \raisebox{1.4cm}{%
      \makebox[\paperwidth][c]{%
        \footnotesize
        Copyright may be transferred without notice, after which this version may no longer be accessible.
      }%
    }%
  }%
}

\begin{abstract}
This full Innovative Practice paper reports on the rapid development and classroom deployment of a Thonny log visualizer built using AI-assisted ``vibe coding'' to make students' programming processes easily visible to teachers. We developed a web application that analyzes log files generated by Thonny (an IDE for Python) and produces interpretable views of students' programming processes. Teachers can upload a log, a ZIP archive, or a folder containing logs for a group or the course; the system parses all logs, generates results per student, and provides student-by-student navigation for reviewing cases. Each student's view includes an interactive activity timeline, a compact session summary, a code-size graph, a programming-process replay, and more. These views support teacher decision-making by enabling the identification of learning-support situations and flagging sessions for academic-integrity clarification. The tool was initially evaluated using logs from previous courses; a February 2026 pilot in an introductory programming course with 160 participants provided teachers’ feedback and informed iterative usability improvements.
\end{abstract}


\section{Introduction}
Introductory programming courses often face challenges related to large enrollments and diverse student backgrounds \cite{luxton2018,omalley2020}. In addition, research has emphasized the importance of supporting students effectively already in the first weeks of study \cite{estey2016,porter2014}. In introductory programming courses, teachers typically assess students' work mainly through final submissions, while much of the underlying programming process remains hidden. Important signals of programming process behavior, such as runs, errors, debugs, pauses, and copy--paste behavior, are often not visible in typical teaching workflows, even though they can provide valuable insights into how students approach tasks, where they struggle, and when they may need support. For example, beginners may make frequent trial-and-error attempts \cite{blikstein2011,hosseini2014}, engage in extensive copy--and--paste behavior \cite{vihavainen2014}, or start executing their programs only late in the process or not at all \cite{carter2017,taveter2025}. Such differences in process behavior may help explain differences in student progress and performance \cite{pereira2020,taveter2025}. Programming-process data has also been found useful for supporting formative feedback. In particular, it can provide teachers with more detailed information about learners' difficulties \cite{ebrahimi2012,yan2019} and enable earlier, more targeted interventions \cite{meierlepp2023b,rocha2023,zhangvizprog2023}.

This limited visibility creates several practical challenges. Teachers may identify struggling students only after substantial time has passed or after a final submission has already been produced. Opportunities for timely formative feedback may be missed because the process leading to the final outcome is not observable. In addition, when plagiarism is suspected, the available evidence may be restricted to the final code rather than the broader programming process. This is particularly problematic in introductory courses, where final programs are often short and similar \cite{modiba2016,ryman2021}. In such cases, source-code comparison alone may be insufficient, especially when students modify or obfuscate copied code \cite{herrera2019,novak2019}.

Although programming-process-based tools have been proposed for feedback and visualization \cite{yan2019,yu2023,zhangvizprog2023}, and related approaches have also been used for plagiarism-related analysis \cite{hart2023,rodriguezrivera2022,shrestha2022}, their classroom use can still be time-consuming, especially when teachers must inspect students' work individually. There is therefore a need for teacher-facing tools that provide interpretable process views, support comparison across students, and help flag cases that require closer inspection. These needs are particularly pronounced in large introductory courses, where several teachers may need to consistently and efficiently review many students' work.

To address these challenges, we developed Thonny Log Visualizer, a web application for teachers that analyzes log files generated by Thonny, an integrated development environment (IDE) for Python \cite{annamaa2015}. The tool produces interpretable views of students' programming processes and supports both single-student inspection and cohort-level analysis through file, ZIP, or folder upload. Rather than serving as a simple log viewer, the tool is intended to support teacher triage, formative feedback, and closer review of cases requiring further inspection. The visualizer was developed rapidly using an AI-assisted ``vibe coding'' workflow \cite{meske2025,pajo2025}, aiming to produce a classroom-ready tool that helps teachers support students. This paper presents the visualizer as a teacher-facing tool for process-informed support, describes its AI-assisted development approach, and reports its validation, iterative refinement, and teacher pilot in an authentic course context. The novelty of the work lies in combining teacher-facing programming-process visualization, cohort-level case prioritization, and lightweight integrity-related indicators into a classroom-ready tool developed and iteratively refined through an AI-assisted workflow.

This paper is structured as follows. Section~II presents background information and related work. Section~III describes the educational context, design goals, and development approach. Section~IV presents the Thonny Log Visualizer. Section~V discusses the validation and evaluation of the tool. Finally, Section~VI provides the conclusion and directions for future work.

\section{Background and Related Work}
\subsection{Programming Process Analysis in Introductory Programming}
Over the past decade, computing education research has increasingly emphasized the importance of analyzing the programming process rather than focusing solely on final programs \cite{edwards2023,taveter2025}. Modern IDEs and educational environments enable the collection of detailed activity logs that capture events such as program runs, error messages, debugging actions, code edits, pauses, file operations, and copy--paste behavior \cite{blikstein2011,leinonen2016,meierkutt2024}. These data provide opportunities to examine how students iteratively construct solutions and how their interaction patterns relate to learning outcomes \cite{pereira2020,leinonen2022,taveter2025}.

Prior research has demonstrated that a wide range of measurable features can be extracted from programming-process logs and used to characterize how students work. These include the number and timing of runs and compilations \cite{hosseini2014,jadud2006,sharma2018}, frequency and repetition of error messages \cite{carter2015,price2020,tabanao2011}, use of debugging tools \cite{watson2013,zhangdebug2023}, and temporal gaps or pauses during programming \cite{leinonen2022,shrestha2022}. Additional indicators, such as proportion of pasted text \cite{vihavainen2014}, number of opened or reused files \cite{blikstein2011}, and patterns of code growth and reduction over time \cite{hosseini2014,shrestha2022,yan2019}, provide further insight into students' problem-solving dynamics. Studies using logs demonstrate that such features can be systematically aggregated and analyzed across sessions to characterize students' programming processes \cite{meier2020,meierlepp2023a,taveter2025}.

Beyond descriptive research analysis, programming-process data has also been leveraged to support formative feedback in introductory programming courses. Log-based feedback interventions have been shown to improve exam performance and reduce task completion time, particularly among beginners \cite{meierlepp2023b}. Adaptive, process-aware feedback mechanisms have also been implemented in block-based environments, where student actions are continuously analyzed to generate context-sensitive guidance during task completion \cite{marwan2020}. Several visualization tools have been developed to make process data easier for learners and instructors to interpret. Replay-based systems allow students to review their own development trajectories and examine metrics such as solving time, number of compilations, and intermediate states \cite{matsuzawa2013}. Other approaches provide visual summaries of code evolution and interaction patterns to support teacher--student discussions about problem-solving strategies and to help instructors identify struggling students \cite{yan2019,shrestha2022}. While pedagogically promising, many of these tools are developed within controlled settings, making it difficult for instructors to adopt and integrate them quickly into everyday classroom practice.

Academic integrity presents an additional motivation for examining programming-process data in introductory programming courses, where assignments are often short and structurally similar, making static source-code comparison vulnerable to superficial modifications such as renaming variables, reordering statements, or minor syntactic changes \cite{novak2019,duric2013}. As a result, researchers have explored behavioral signals embedded in programming logs as complementary evidence. Process data provides indicators that are difficult to manipulate retrospectively, including sudden increases in code size due to paste events, minimal intermediate edits before arriving at a correct solution, highly linear development trajectories, or unusual consistency in timing patterns \cite{hellas2017,karnalim2019}. Keystroke dynamics and digraph timing have been investigated as potential identifying features \cite{byun2020,longi2015}, and log-based comparisons of command usage patterns have also been proposed \cite{schneider2018}. Further analyses suggest that process-level indicators can complement source-code similarity by helping teachers focus attention on cases that warrant closer inspection \cite{meierlepp2021,meierkutt2024}.

\subsection{Vibe Coding}
Recent advances in generative AI have introduced new paradigms for software development, often described as ``vibe coding,'' in which applications are developed primarily through natural-language prompting and iterative refinement of AI-generated outputs \cite{pajo2025}. Conceptual work frames vibe coding as a distinct shift in how developer intent is mediated: rather than translating goals into deterministic code instructions, developers increasingly articulate high-level intent conversationally, with AI systems inferring and implementing solutions probabilistically \cite{meske2025}. This redistribution of cognitive labor shifts emphasis from low-level implementation to higher-level orchestration and evaluation, while also introducing risks, such as incomplete understanding of AI-generated implementations and gaps in responsibility \cite{meske2025}.

Empirical research in computing education has begun to explore how this paradigm manifests in learning contexts. Studies of educational hackathons show that novice and mixed-experience teams can produce functional prototypes within short time frames using vibe coding, while also facing challenges such as uneven code quality and limited exploration before committing to a solution \cite{gama2026}. Observational studies further indicate that students using vibe coding platforms primarily engage in prototype testing and debugging through prompts, with relatively limited direct modification of generated code \cite{geng2025}. Differences by experience level have also been observed: more advanced students tend to formulate code-focused prompts and incorporate contextual reasoning, whereas less experienced students rely more heavily on high-level or low-context instructions \cite{geng2025}. Related research on generative AI in programming similarly indicates that effective prompting and verification depend on prior programming knowledge \cite{kazemitabaar2024,prather2024}. Educational research has also begun to position vibe coding within programming pedagogy. In this framing, AI is conceptualized not merely as a tool but as a potential teammate within project-based learning environments, with preliminary evidence suggesting perceived gains in productivity and engagement when students collaborate with AI systems \cite{kusper2025}. Such perspectives emphasize flow, iterative dialogue, and evolving intent as central characteristics of AI-mediated development in classroom settings.

Research on professional developers provides a complementary perspective. Studies of extended vibe coding sessions suggest that the need for programming expertise is not eliminated but reconfigured: experienced developers engage in rapid evaluation, selective editing, and strategic transitions between AI-driven generation and manual intervention \cite{sarkar2025}. Rather than replacing programming knowledge, vibe coding shifts its emphasis toward contextual reasoning, system-level thinking, and critical oversight of generated artifacts.

While current research focuses on the use of vibe coding by novice students and expert developers, less attention has been paid to instructors and researchers, who often occupy an intermediate position: they possess substantial programming knowledge but typically operate under different constraints, responsibilities, and time pressures than professional software engineers. In particular, there is limited discussion of how AI-assisted development can empower educators to move from conceptual idea to classroom-deployed educational tools within days.

\section{Context, Design Goals, and Development Approach}
The visualizer was developed for use in introductory programming courses where students work in Python using the Thonny integrated development environment. Large courses with several instructors teaching different groups were given particular consideration. In this context, teachers need to review both individual student work and larger sets of student logs across a course. The practical need was therefore in two directions: first, to support closer inspection of individual programming sessions when a student appears to struggle; and second, to support efficient case comparison across many students, including situations where process evidence may help clarify possible plagiarism concerns.

Based on this context, we formulated the following design goals:

\textbf{DG1. Make students' programming processes visible to teachers.} The tool should provide interpretable views that summarize and visualize the necessary signals of programming behavior.

\textbf{DG2. Support teacher decision-making through various workflows.} The application should support both quick inspection of individual cases and efficient review of larger sets of student logs.

\textbf{DG3. Support process-informed formative feedback.} The visualizer should help teachers identify situations in which students may need support, such as repeated run-error cycles with limited progress, a late start of the program execution, or unusually long periods of inactivity.

\textbf{DG4. Support clarification of academic-integrity concerns without automating judgment.} The tool should surface signals such as pasted-text proportion, paste source, similarity, and duplicate function names, while leaving interpretation to teachers. The application must also allow replay of the programming process for a more detailed review of cases.

To respond quickly to this instructional need, the visualizer was developed through an AI-assisted workflow, enabling rapid implementation and refinement of teacher-facing functionality. The main development environment was Lovable \cite{lovable2026}, which was used to generate and modify the web application from natural-language prompts. Initial prompts were drafted with the support of the large language model ChatGPT 5.2 and were then reviewed and revised by the first author before submission to Lovable. Prompts to ChatGPT typically included the instruction ``make a prompt for Lovable'' followed by a plain-language description of what needed to be implemented or fixed. ChatGPT-generated prompts were typically structured to begin with a similar opening sentence: ``The app already exists. Fix ONLY [X]. Do not change analysis logic, calculations, or UI content---only [X].'' This was followed by the sections ``Problem,'' ``Required behavior,'' ``Implementation details,'' and ``Acceptance criteria.'' In some cases, prompts were submitted to Lovable without the ``Implementation details'' section to avoid overly prescriptive guidance. Development proceeded incrementally: after the initial tool, which focused on some student-level views (activity timeline, code size over time, and session summary) based on a single log, subsequent work was carried out feature by feature. Each newly implemented view, component, or feature was tested before the next one was developed. Throughout the process, the tool was repeatedly inspected and tested using diverse Thonny log files and refined through further prompting and revision. Typically, a single prompt could quickly produce substantial functionality, but testing it required more time and additional prompts. In some cases, extra effort was needed to account for differences in log structure, such as those between logs generated by older and newer versions of Thonny or by computers with different operating systems. A total of 207 Lovable prompts and 403 Lovable credits were needed to create the application, test different views, components, and functions, and implement changes and improvements that emerged during piloting. Compared to other views, developing the programming-process replay required the largest number of prompts (31): to make it work reliably and consistently, ensure that information across tabs and logs is displayed correctly, allow the user to inspect points of interest in more detail, and account for edge cases.

The process can be summarized as follows: the first author defined the instructional needs, selected the features and views to include, evaluated whether the generated functionality met the intended use, and wrote the next prompts accordingly. In this way, AI assistance lowered the technical barrier and accelerated implementation, while pedagogical reasoning, feature selection, and judgments about usefulness remained under human control.

\section{New Programming Process Visualizer}
\subsection{Overview and Input Workflows}
Thonny Log Visualizer is a web application that analyzes log files generated by Thonny, an integrated development environment (IDE) for Python, and produces interpretable views of students' programming processes. Rather than implementing isolated metrics, the visualizer translates prior findings on programming-process data into a teacher-facing set of views intended to support rapid orientation, closer inspection, and comparison across students. The application supports several input workflows. For quick inspection, a teacher can upload a single student log and immediately open a session dashboard. For broader review, the teacher can upload either a ZIP archive or an entire folder containing multiple students' logs. In the latter case, the system parses all logs, groups them by student, generates results per student, and supports student-by-student navigation for reviewing cases. This design enables both individual-case analysis and cohort-level use. A teacher may therefore inspect one student in depth, or start from a larger group and then move to selected students whose metrics or warning indicators suggest that closer review is needed.

\subsection{Student-Level Dashboard}
For each student, the application provides a session dashboard that integrates several complementary views. The choice of views and indicators was informed by earlier research on programming-process data. Prior work has shown the value of replaying programming activity and examining metrics such as solving time, runs, and compilation-related events \cite{matsuzawa2013}, while Shrestha et al.~\cite{shrestha2022} highlighted the usefulness of pauses and of visualizing the programming process. Research has also shown that errors and debugging-related behavior can be important for understanding student performance and difficulties \cite{watson2013,zhangdebug2023}. Indicators related to pasted text and code modification were informed by prior work showing that process-level plagiarism analysis can benefit from keystroke and log data \cite{hellas2017,hart2023}, while views of inserted and deleted code can provide useful summaries of how a solution evolves over time \cite{rodriguezrivera2022}. The selection of indicators was further informed by earlier work on the timing of first program execution \cite{taveter2025} and on persistent symbol-writing order \cite{meierlepp2021}. Based on these considerations, the main views include an activity timeline, a compact session summary, a code-size graph, and a programming-process replay with an event feed, complemented by a paste-events view and additional indicators related to final code, similarity, first-run progress, and pair typing patterns.

\begin{figure}[t]
\centering
\includegraphics[width=\columnwidth]{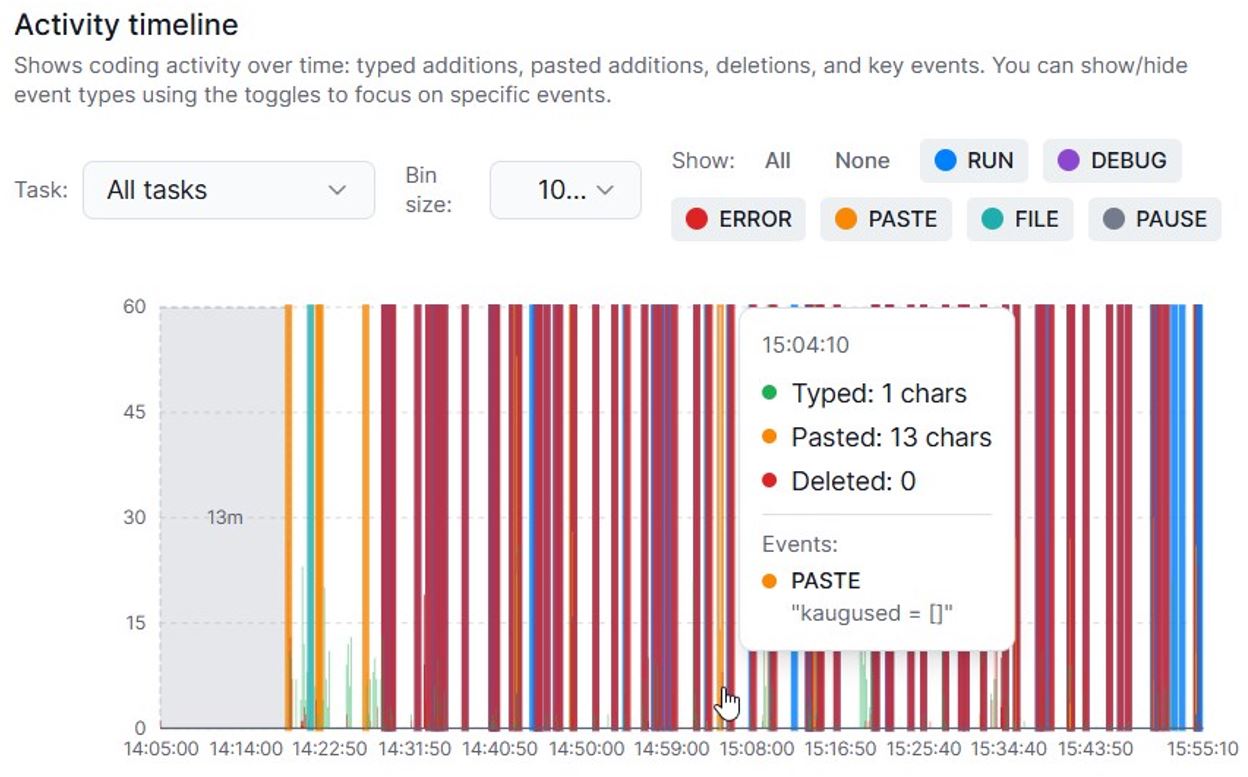}
\caption{Activity timeline.}
\label{fig:timeline}
\end{figure}

One core component is an interactive activity timeline showing the temporal distribution of runs, debugs, errors, pastes, file actions, pauses, and typed/deleted text (Fig.~\ref{fig:timeline}). Teachers can filter events by type, adjust the time bin size, and inspect event details directly from the timeline. This supports rapid recognition of temporal patterns such as dense run-error sequences, bursts of activity, or longer pauses. The dashboard also includes a compact session summary containing aggregate indicators such as session duration, active time, pauses, inserted and deleted characters, net change, runs, debugs, errors, pastes, and files created. These indicators provide immediate orientation before more detailed analysis.

\begin{figure}[t]
\centering
\includegraphics[width=\columnwidth]{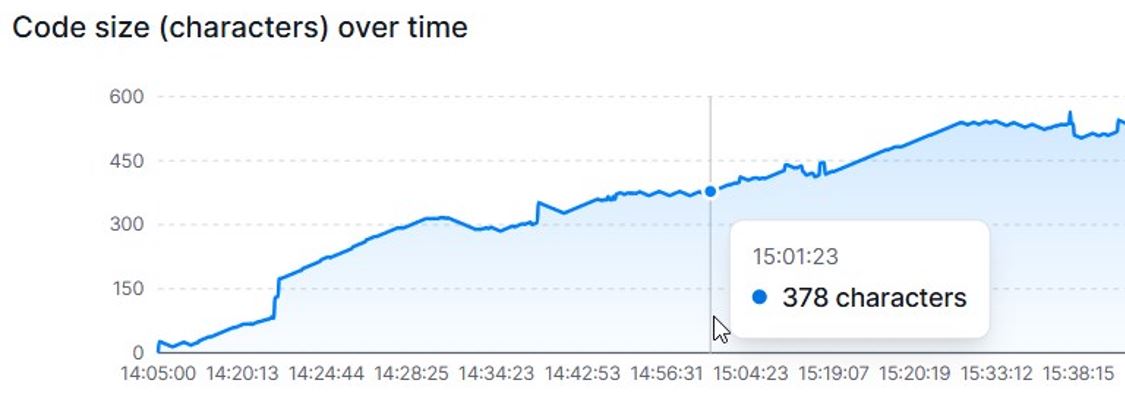}
\caption{Code size graph.}
\label{fig:codesize}
\end{figure}

A code size graph shows how code size changes over time (Fig.~\ref{fig:codesize}). This can help teachers identify gradual progress and abrupt increases that may warrant closer inspection.

\begin{figure}[t]
\centering
\includegraphics[width=\columnwidth]{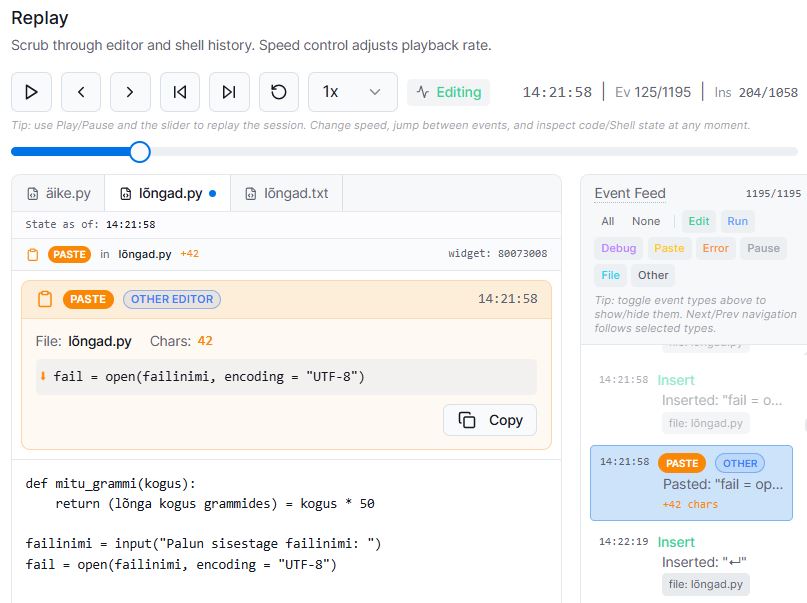}
\caption{Programming-process replayer.}
\label{fig:replayer}
\end{figure}

To support fine-grained examination of process dynamics, the system also provides programming-process replay, accompanied by an event feed listing events in sequence (Fig.~\ref{fig:replayer}). Replay enables step-by-step reconstruction of the programming process and explicitly represents inactivity so that pauses are not confused with replay failure. Users can also change the replay speed, use the slider to jump to another point in the review, filter events in the event feed by type, jump to the next marker event, and then continue replaying.

\begin{figure}[t]
\centering
\includegraphics[width=\columnwidth]{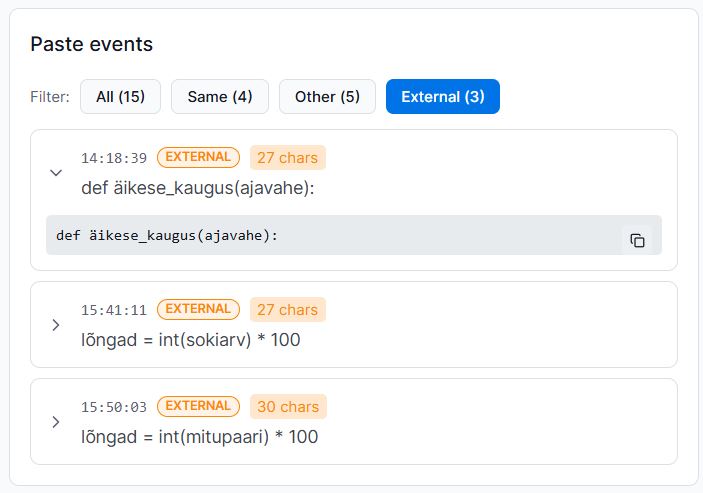}
\caption{Paste events.}
\label{fig:pasteevents}
\end{figure}

In addition to the main dashboard, the application provides several specialized views. A ``Paste events'' section distinguishes between pastes originating from the same editor, another tab, and outside the IDE, and presents pasted snippets with timestamps and size information (Fig.~\ref{fig:pasteevents}). This supports a more nuanced interpretation of pasted-text proportion than a single aggregate value alone.

\begin{figure}[t]
\centering
\includegraphics[width=\columnwidth]{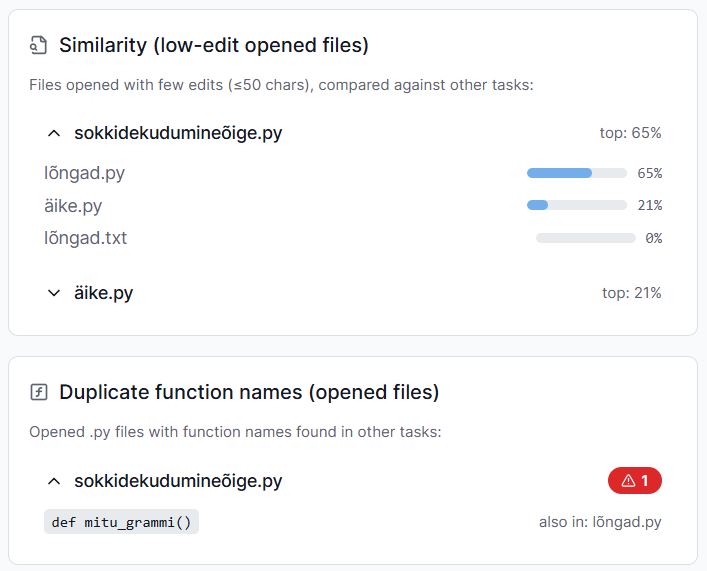}
\caption{Similarity and duplicate function-name indicators.}
\label{fig:similarity}
\end{figure}

The application also presents the final code and lightweight integrity-related indicators, including similarity information for low-edit opened files and duplicate function names across opened files (Fig.~\ref{fig:similarity}). These features are intended as warning signals rather than automated determinations.

Two further views address process structure more directly. The ``Code present by first run'' view estimates how much code was already present at the time of first execution and how early that first run occurred relative to the total active time. The pair typing patterns view analyzes the order in which paired symbols, such as brackets and quotation marks, are entered, distinguishing patterns such as open $\rightarrow$ text $\rightarrow$ close and open $\rightarrow$ close $\rightarrow$ text. These indicators are not interpreted in isolation; they may provide additional context when considered alongside paste behavior and other process information.

\subsection{Cohort-Level Comparison Views}
\begin{figure*}[t]
\centering
\includegraphics[width=0.95\textwidth]{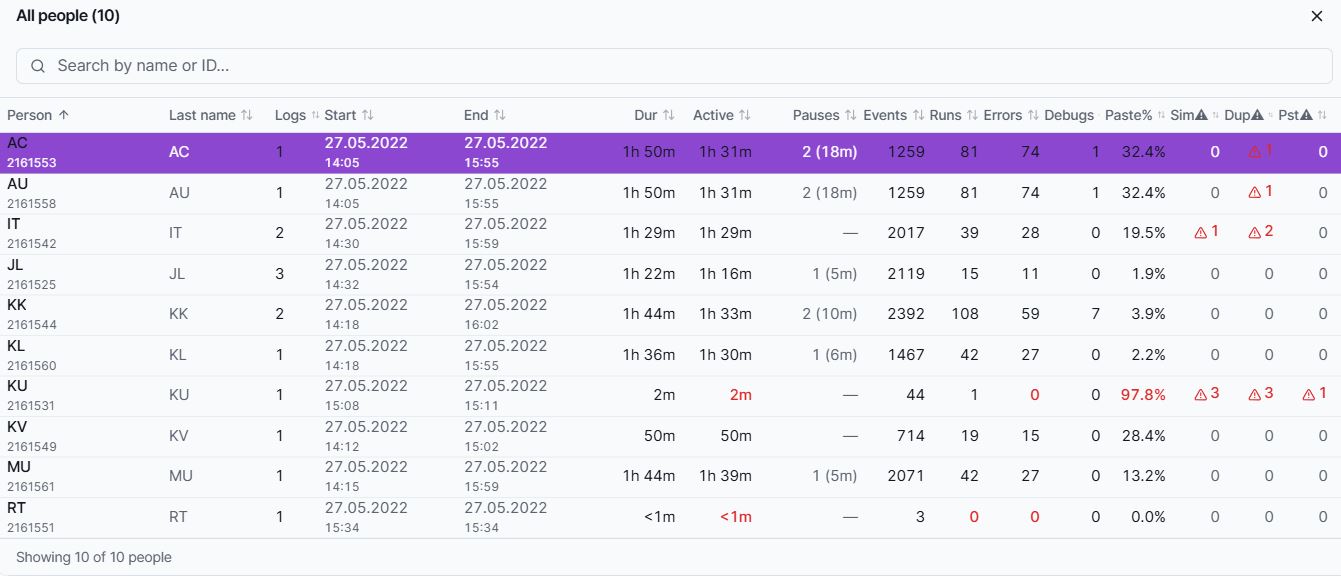}
\caption{``All people'' view for cohort-level comparison.}
\label{fig:allpeople}
\end{figure*}

To support analysis beyond individual students, the system includes several cohort-level comparison views. The ``All people'' view provides a sortable table of students with summary measures, including the number of logs, start and end times, events, duration, active time, pauses, pasted-text proportion, and warning indicators for similarity, duplicate function names, and paste share (Fig.~\ref{fig:allpeople}). This enables teachers to identify outliers and prioritize cases for closer inspection. Clicking a selected record opens the student-level view for that student.

The ``Compare pair typing patterns'' view summarizes pair-typing distributions across the cohort and groups students according to their proportion of alternative typing-order patterns. Similarly, the ``Compare first run progress'' view compares how much code students had written by their first run and when that first run occurred. These cohort-level views allow individual cases to be interpreted relative to others' behavior rather than in isolation.

\section{Validation and Evaluation}
\subsection{Validation with Diverse Thonny Logs}
The application was validated iteratively during development through repeated testing with diverse Thonny logs. Validation used more than 20 different ZIP archives or folders, each containing a set of logs from 5 to 87 students. These logs varied in length, number of events, file structure, paste behavior, and session characteristics, allowing the authors to check parsing accuracy, consistency of computed indicators, and alignment between the underlying logs and the visual representations shown in the application. Validation also focused on edge cases (e.g., long inactivity periods, multi-file sessions, and frequent paste events) to ensure robust behavior across typical course logs. This stage was intended to establish that the visualizations and summary indicators remained aligned with the underlying log data across varied session structures and edge cases before classroom use. The application was also tested with logs from cases already known to involve plagiarism. All of these cases were flagged by the application (Fig. 6). Examining the flagged entities in more detail through the student-level views enabled identification of the relevant cases. These results suggest that the flagging mechanisms were sensitive enough to surface relevant cases for follow-up inspection, while still requiring student-level review rather than replacing teacher judgment.

\subsection{Developer-Led Evaluation and Iterative Refinements}
The evaluation focused on convenience, understandability, and whether the views supported meaningful teacher interpretation. The iterative review led to several refinements in both functionality and usability. First, handling of larger ZIP uploads was improved to support more convenient student-by-student review when many logs are analyzed together. Second, the replay was modified to explicitly show inactivity, reducing ambiguity during longer pauses in student work; the replay was also improved with the option to filter events in the event feed by type and jump to that point, allowing teachers to more easily find the part of the programming process that needs review. Third, paste metrics were refined to distinguish between pastes from the same editor, another tab, and outside the IDE. Fourth, to give users greater convenience when using different input formats, folder input was added alongside single-log and ZIP inputs. Fifth, to diversify the analysis and give more information for formative feedback, a view was added that shows how much code students wrote by their first run and when that first run occurred. Sixth, to improve understandability, user-facing descriptions and help text were added for sections, columns, terms, and buttons, with many of these explanations appearing as hover text when the cursor is placed over the relevant interface element.

\subsection{Teacher Pilot in an Introductory Programming Course}
In the final stage of development, teachers used the tool in an introductory programming course and provided feedback that helped improve usability. The pilot was conducted in the ``Introduction to Programming'' course, which enrolled 160 students, and the tool was used by five teachers. During the pilot, teachers reviewed logs using both single-student and cohort workflows (ZIP/folder upload) as part of routine course support and case review. Feedback was gathered through informal teacher-led reviews of the tool and written comments during the course. Overall feedback highlighted the value of the visualizer for gaining rapid, process-level insight. Teachers particularly appreciated the initial visual overview of different activity types and the clearly marked pauses, which helped them quickly orient themselves in a session. They also noted that the tool enables detailed inspection of how a student actually solved a task, and that having the final solution/code clearly visible alongside process views is useful. The tool also helped teachers identify cases in which solutions were largely copied. In addition, it made it easier to detect cases in which a student had submitted an incorrect or incomplete log file. Building on this feedback, the pilot led to several concrete usability improvements that were implemented iteratively. Corrections were typically completed within a few days of receiving feedback, while additional improvements were implemented after all participating teachers had provided their input. The following changes represent the main updates resulting from the pilot.

\textit{Corrections and Performance Fixes.} Teachers reported that switching between students could be slow in some cases; performance was improved to support faster student-by-student review. Teachers also reported that the replay focus jumped to the lowest entry in the event feed, potentially hiding the top navigation bar on smaller screens; this behavior was adjusted to improve navigation and visibility. In addition, the replay slider was inconsistent (``jumpy''), so it was refined to provide a smoother experience.

\textit{Enhancements.} Additional replay speed options were requested and implemented by adding 15$\times$ and 30$\times$ speeds. Teachers also suggested adding a visual indicator in the ``All people'' view by coloring the solving time red when it is very short; this was implemented by coloring the text red when it is below 10 minutes. It was suggested to add the ability to select a time window to control which logs are displayed and analyzed; this has not yet been implemented. Based on feedback, information was added to the main view indicating which folders contain files not suitable for analysis. The folder/ZIP file name was also added to the visible area.

\section{Conclusion}
This paper presented Thonny Log Visualizer, a teacher-facing web application that analyzes IDE activity logs and produces interpretable views of students' programming processes. Developed rapidly through AI-assisted ``vibe coding,'' the tool supports both individual and cohort-level review through student dashboards, replay, paste analysis, and comparison views. The work shows that programming-process visualization can be made classroom-ready with a relatively low development barrier and can support formative feedback, earlier identification of struggling students, and clarification of academic-integrity concerns. In practice, the tool supports teacher triage by combining student-level dashboards with cohort-level views that help prioritize cases for closer inspection. A pilot with five teachers in a 160-student introductory course highlighted the value of the programming process visualization and led to iterative usability improvements. However, the evaluation was conducted within a single course context, limiting generalizability. In addition, the interpretable views of students' programming processes are intended to support follow-up discussion and manual review, rather than to serve as automated judgments. A limitation is that the visualizer depends on the course requirement that students complete the mandatory tasks in Thonny and submit the corresponding log files. For computing education practice, the results suggest that process-oriented teacher tools can be integrated into routine course support without requiring fully automated judgment, provided that the underlying logging workflow is available. Future work includes adding time-window filtering to displayed logs, extending cohort-level analytics, and further studying how teachers use process signals in instructional decision-making. More broadly, the work illustrates how prior learning-analytics research can be operationalized into lightweight classroom tools and suggests that AI-assisted development can help teachers and education-focused teams create and iterate on specialized instructional tools in the LLM era, provided that validation and careful interpretation remain central.
\\
\section*{Acknowledgment}

This work was supported by the Estonian Research Council grant ``Developing human-centric digital solutions'' (TEMTA120).

\balance

\end{document}